\documentclass[runningheads]{llncs}
\usepackage{graphicx}
\usepackage{amsmath}
\begin{document}
\title{JBFnet - Low Dose CT Denoising by Trainable Joint Bilateral Filtering}
\author{Mayank Patwari\inst{1, 2}\orcidID{0000-0001-5197-5260} \and
Ralf Gutjahr\inst{2} \and
Rainer Raupach\inst{2} \and
Andreas Maier\inst{1}}
%
%index{Patwari, Mayank}
%index{Gutjahr, Ralf}
%index{Raupach, Rainer}
%index{Maier, Andreas}

\authorrunning{M. Patwari et al.}
\institute{Pattern Recognition Lab, Friedrich-Alexander Universität Erlangen-Nürnberg (FAU), Erlangen, Germany \and
Siemens Healthcare GmbH, Forchheim, Germany \\
\email{mayank.patwari@fau.de}\\
}
\maketitle              % typeset the header of the contribution
\begin{abstract}
Deep neural networks have shown great success in low dose CT denoising. However, most of these deep neural networks have several hundred thousand trainable parameters. This, combined with the inherent non-linearity of the neural network, makes the deep neural network difficult to understand with low accountability. In this study we introduce JBFnet, a neural network for low dose CT denoising. The architecture of JBFnet implements iterative bilateral filtering. The filter functions of the Joint Bilateral Filter (JBF) are learned via shallow convolutional networks. The guidance image is estimated by a deep neural network. JBFnet is split into four filtering blocks, each of which performs Joint Bilateral Filtering. Each JBF block consists of 112 trainable parameters, making the noise removal process comprehendable. The Noise Map (NM) is added after filtering to preserve high level features. We train JBFnet with the data from the body scans of 10 patients, and test it on the AAPM low dose CT Grand Challenge dataset. We compare JBFnet with state-of-the-art deep learning networks. JBFnet outperforms CPCE3D, GAN and deep GFnet on the test dataset in terms of noise removal while preserving structures. We conduct several ablation studies to test the performance of our network architecture and training method. Our current setup achieves the best performance, while still maintaining behavioural accountability.

\end{abstract}
\begin{figure}[t]
	\includegraphics[width=\linewidth]{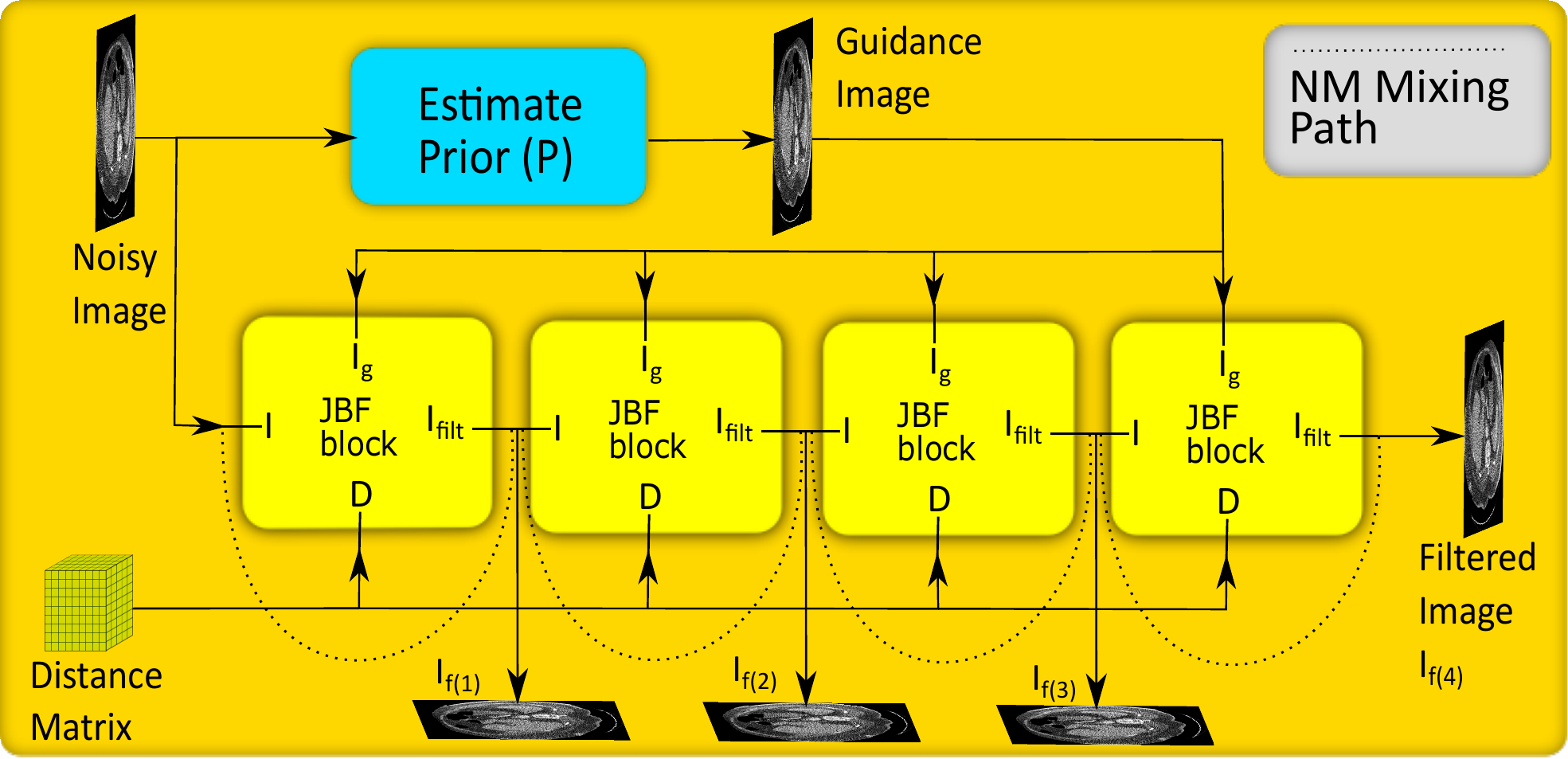}
	\caption{The general architecture of JBFnet. There is a prior estimator and 4 JBF blocks for filtering. The NM mixing paths are shown in the diagram. The intermediate outputs of JBFnet are also shown.}
	\label{JBFnetArch}
\end{figure}

\section{Background}
Reducing the radiation dose in clinical CT is extremely desirable to improve patient safety. However, lowering the radiation dose applied results in a higher amount of quantum noise in the reconstructed image, which can obscure crucial diagnostic information \cite{Oppelt2005a}. Therefore, it is necessary to reduce the noise content while preserving all the structural information present in the image. So far, this is achieved through non-linear filtering \cite{Manduca2009,Manhart2014} and iterative reconstruction \cite{Gilbert1972,Hounsfield1973,Kaczmarz1937}. Most CT manufacturers use iterative reconstruction techniques in their latest scanners \cite{Angel2012,Ramirez-Giraldo2015}. 

In the past decade, deep learning methods \cite{Goodfellow2014} have shown great success in a variety of image processing tasks such as classification, segmentation, and domain translation \cite{Lecun2015}. Deep learning methods have also been applied to clinical tasks, with varying degrees of success \cite{Nie2017,Wolternik2017}. One such problem where great success has been achieved is low dose CT denoising \cite{Fan2019,Li2020,Maier2015,Shan2018,Wolterink2017,Yang2018,Zhang2017a}. Deep learning denoising methods outperform commercially used iterative reconstruction methods \cite{Shan2019}. 

Deep learning methods applied to the problem of low dose CT denoising are usually based on the concept of deep convolutional neural networks (CNN). Such CNNs have hundreds of thousands of trainable parameters. This makes the behaviour of such networks difficult to comprehend, which aggravates the difficulty of using such methods in regulated settings. This is why, despite their success, deep CNNs have only sparingly been applied in the clinic. It has been shown that including the knowledge of physical operations into a neural network can increase the performance of the neural network \cite{Maier2018,Maier2019,Syben2017}. This also helps to comprehend the behaviour of the network. 

In this work, we introduce JBFnet, a CNN designed for low dose CT denoising. JBFnet implements iterative Joint Bilateral Filtering within its architecture. A deep CNN was used to generate the guidance image for the JBF. Shallow 2 layered CNNs were used to calculate the intensity range and spatial domain functions for the JBF. JBFnet has 4 sequential filtering blocks, to implement iterative filtering. Each block learns different range and domain filters. A portion of the NM is mixed back in after filtering, which helps preserve details. We demonstrate that JBFnet significantly outperforms denoising networks with a higher number of trainable parameters, while maintaining behavioural accountability.

\section{Methods}
\subsection{Trainable Joint Bilateral Filtering}
JBF takes into account the differences in both the pixel coordinates and the pixel intensities. To achieve this, the JBF has separate functions to estimate the appropriate filtering kernels in both the spatial domain and the intensity ranges. As opposed to bilateral filtering, the JBF uses a guidance image to estimate the intensity range kernel. The operation of the JBF is defined by the following equation:

\begin{equation}
\label{JBFeqn}
I_{f} (x)=  \frac{\sum_{o\epsilon N(x)} I_n(o) B_w (I_g, x, o)}{ \sum_{o\epsilon N(x)} B_w (I_g, x, o)} \; ; \; B_w(I_g, x, o) =  G(x-o)  F(I_g (x)- I_g (o))
\end{equation} 

\noindent where $I_n$ is the noisy image, $I_f$ is the filtered image, $x$ is the spatial coordinate, $N(x)$ is the neighborhood of $x$ taken into account for the filtering operation, $I_g$ is the guidance image, $F$ is the filtering function for estimating the range kernel, and $G$ is the filtering function for estimating the domain kernel.

We assume that $G$ and $F$ are defined by parameters $W_g$ and $W_f$ respectively. Similarly, we assume that the guidance image $I_g$ is estimated from the noisy image $I_n$ using a function $P$, whose parameters are defined by $W_p$. Therefore, we can rewrite Equation \ref{JBFeqn} as:
\begin{equation}
\label{JBFeqnWeighted}
I_{f} (x)=  \frac{\sum_{o\epsilon N(x)} I_n(o)G(x-o;W_g) F[P(I_n (x);W_p) - P(I_n (o);W_p); W_f]}{\sum_{o\epsilon N(x)} G(x-o;W_g) F[P(I_n (x);W_p )- P(I_n (o);W_p); W_f]}
\end{equation}

\begin{figure}[t]
	\includegraphics[width=\linewidth]{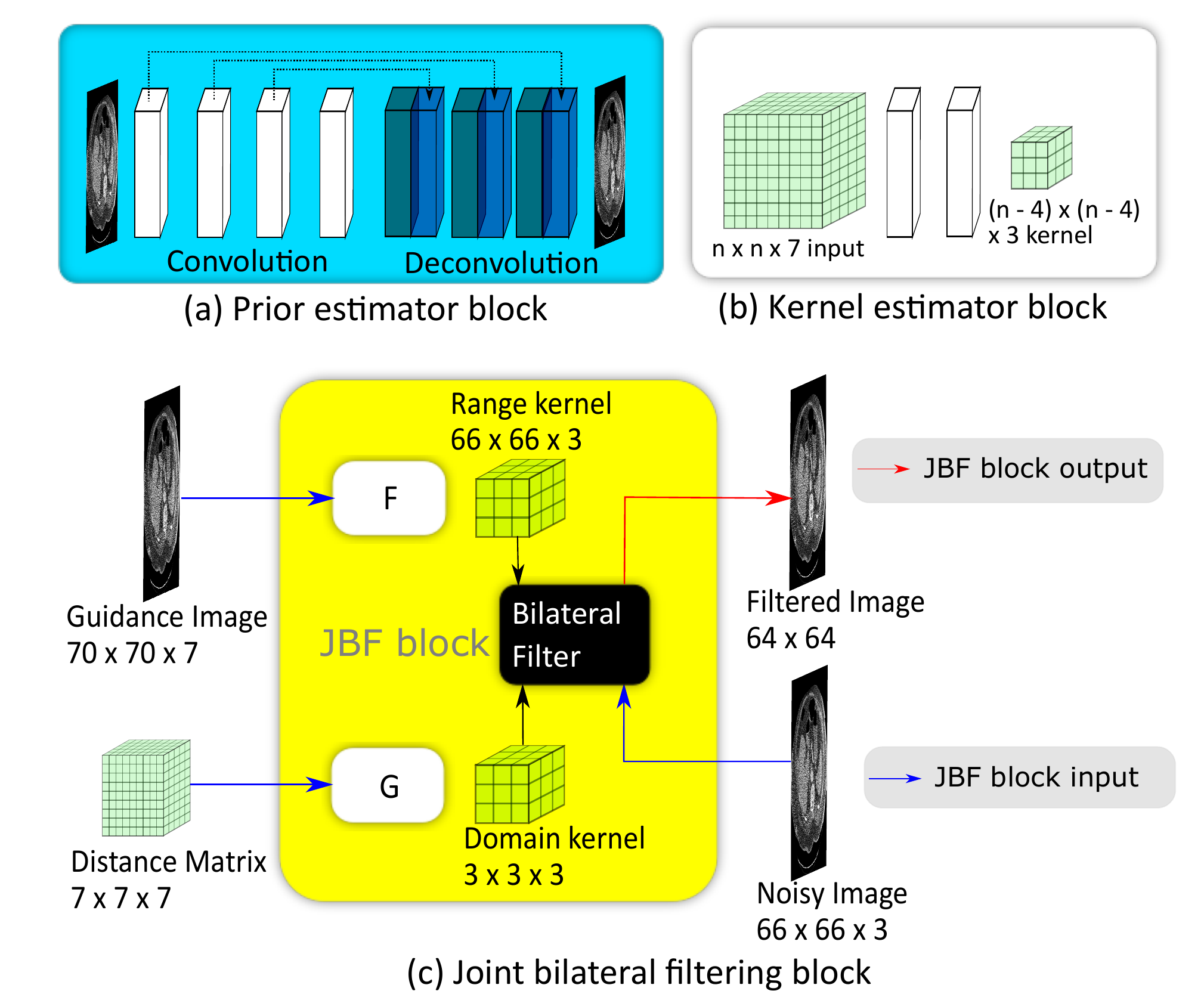}
	\caption{The architecture of (a) the prior estimator network $P$, (b) the kernel estimating functions $G$ and $F$ and  (c) the functioning of the JBF block. Fig. \ref{JBFnetArch} shows how to combine the components to create JBFnet.}
	\label{JBFnetComp}
\end{figure}

\noindent Since the values of $W_f$ act on the guidance image $I_g$ which is generated by $W_p$, we can rewrite $W_f$ as a function of $W_p$. The chain rule can be applied to get the gradient of $W_p$. The least squares problem to optimize our parameters is given by:
\begin{equation}
\label{JBFloss}
\begin{split}
&L(W_g, W_f(W_p))= \\ &\frac{1}{2}||\frac{\sum_{o\epsilon N(x)} I_n(o)G(x-o;W_g) F[P(I_n (x);W_p) - P(I_n (o);W_p); W_f]}{\sum_{o\epsilon N(x)} G(x-o;W_g) F[P(I_n (x);W_p )- P(I_n (o);W_p); W_f]} - I_{f} (x)||^2
\end{split}
\end{equation}
%\begin{equation}
%\label{gradFn}
%dL = \frac{\partial L}{\partial W_g} + \frac{\partial L}{\partial W_p} \implies \frac{\partial L}{\partial W_g} + \frac{\partial L}{\partial W_f} \frac{\partial W_f}{\partial W_p}
%\end{equation}

\noindent The gradients of $W_g$ and $W_p$ ( $\frac{\partial L}{\partial W_g}$ and $\frac{\partial L}{\partial W_f} \frac{\partial W_f}{\partial W_p}$ respectively) can be calculated using backpropogation and the parameters can by updated by standard gradient descent.

\begin{figure}[t]
	\includegraphics[width=\linewidth]{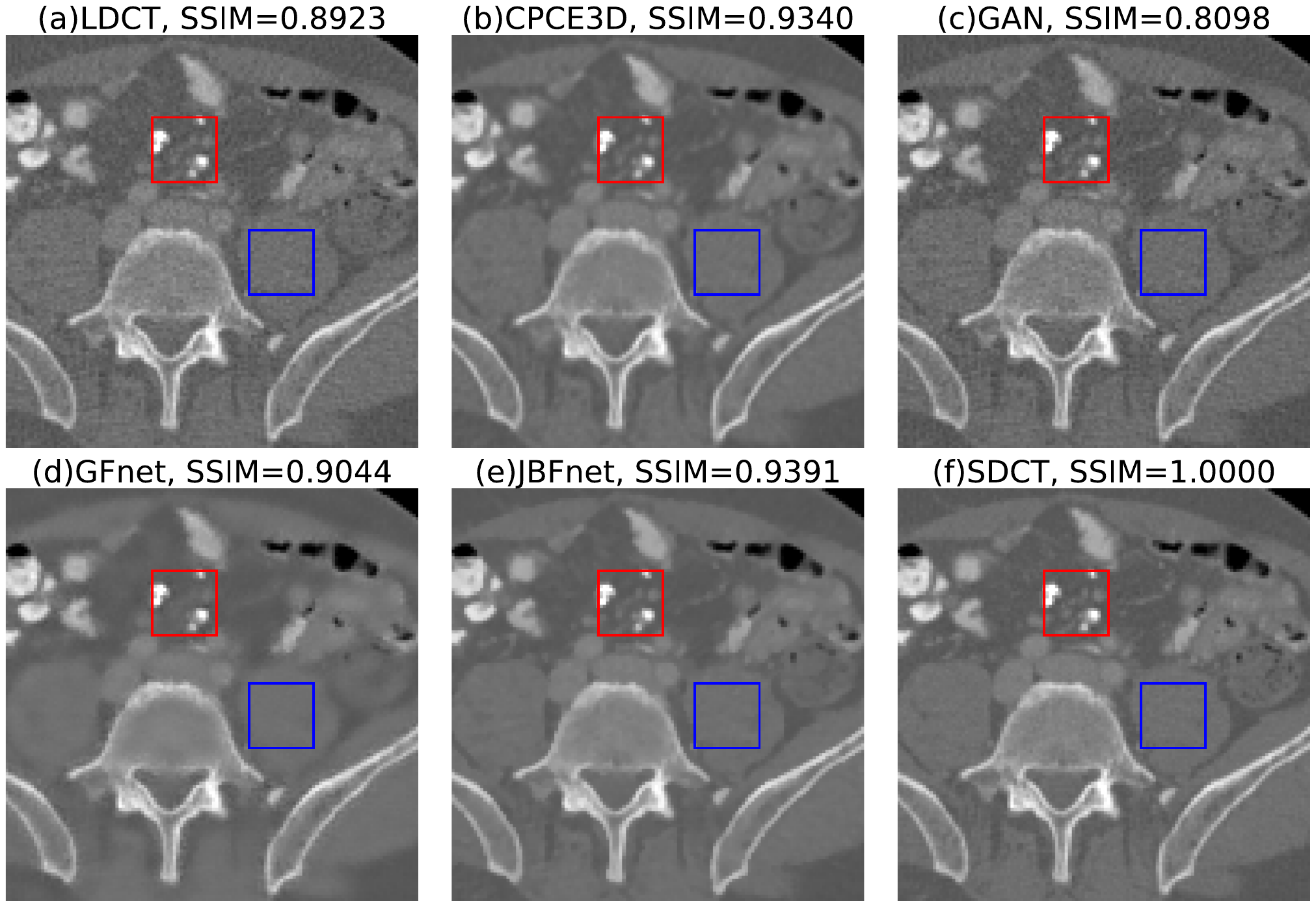}
	\caption{We display an example of (a) low dose CT, denoised by (b) CPCE3D (c) GAN, (d) GFnet (e) JBFnet, and finally (f) the standard dose CT. The red square indicates some small low contrast features. The blue square indicates a homogenous noisy patch. SSIM scores are displayed in the captions. JBFnet achieves the best performance (SSIM = 0.9391). Images are displayed with a window of [-800, 1200].}
	\label{sota}
\end{figure}

\subsection{Network Architecture}

\subsubsection{Prior Estimator Block}
The guidance image is estimated from the noisy image $I_n$ using a function $P$ with parameters $W_p$. We represent this function with a deep neural network inspired by CPCE3D \cite{Shan2018}. We have a hybrid 2D/3D network with 4 convolutional layers and 4 deconvolutional layers. Each layer has 32 filters. The filters in the convolutional layers are of size 3 $\times$ 3 $\times$ 3. The filters in the deconvolutional layers are of size 3 $\times$ 3. Each layer is followed by a leaky ReLU activation function \cite{Maas2013}. $P$ takes an input of size $n\times n \times$ 15, and returns an output of size $n \times n \times$ 7. A structural diagram is present in Fig. \ref{JBFnetComp}(a).

\subsubsection{JBF Block}
We introduce a novel denoising block called the JBF block. JBFnet performs filtering through the use of JBF blocks. Each JBF block contains two filtering functions, $G$ and $F$, with parameters $W_g$ and $W_f$ respectively. We represent each of these functions  with a shallow convolutional network of 2 layers. Each layer has a single convolutional filter of size 3 $\times$ 3 $\times$ 3. Each layer is followed by a ReLU activation function \cite{Maas2013}. No padding is applied, shrinking an input of $n \times n \times$ 7 to $(n - 4) \times (n - 4) \times$ 3 (Fig. \ref{JBFnetComp}(b)). The JBF block then executes a 3 $\times$ 3 $\times$ 3 standard bilateral filtering operation with the outputs of $F$ and $G$. Each JBF block is followed by a ReLU activation function. JBFnet contains four consecutive JBF blocks (Fig. \ref{JBFnetArch}).

\subsubsection{Mixing in the Noise Map}
After filtering with the JBF block, there is the possibilty that some important details may have been filtered out. To rectify this, we mixed an amount of the NM back into the filtered output (Fig. \ref{JBFnetArch}). The NM was estimated by subtracting the output of the JBF block from the input. We mix the NM by a weighted addition to the output of the JBF block. The weights are determined for each pixel by a 3 $\times$ 3 convolution of the NM.

\subsubsection{Loss Functions}
We optimize the values of the parameters using two loss functions. We utilize the mean squared error loss and the edge filtration loss \cite{Shan2019}. The two loss functions are given by the following equations:
\begin{equation}
\label{mseloss}
\begin{split}
MSE Loss (I_1, I_2) = \frac{1}{n} ||I_1 - I_2||^2  ;EF Loss (I_1, I_2) = \frac{1}{n} ||I'_1 - I'_2||^2
\end{split}
\end{equation}

\noindent $I'_1$ and $I'_2$ were computed using 3 $\times$ 3 $\times$ 3 Sobel filters on $I_1$ and $I_2$ respectively. JBFnet outputs the estimated guidance image $I_g$, and the outputs of the four JBF blocks, $I_{f(1)}$,$I_{f(2)}$, $I_{f(3)}$ and $I_{f(4)}$ (Fig. \ref{JBFnetArch}). We indicate the reference image as $I$. Then, our overall loss function is:
\begin{equation}
\label{trainLoss}
\begin{split}
\lambda_1 [ MSELoss(I_{f(4)}, I) + 0.1 * EFLoss(I_{f(4)}, I) ] + \lambda_2 [ MSELoss(I_g, I) ] +\\ \lambda_3 [ \sum_{i=1, 2, 3} MSELoss(I_{f(i)}, I) + 0.1 * EFLoss(I_{f(i)}, I) ]
\end{split}
\end{equation}
where the $\lambda$ values are the balancing weights of the loss terms in Equation \ref{trainLoss}.

\begin{figure}[t]
	\includegraphics[width=\linewidth]{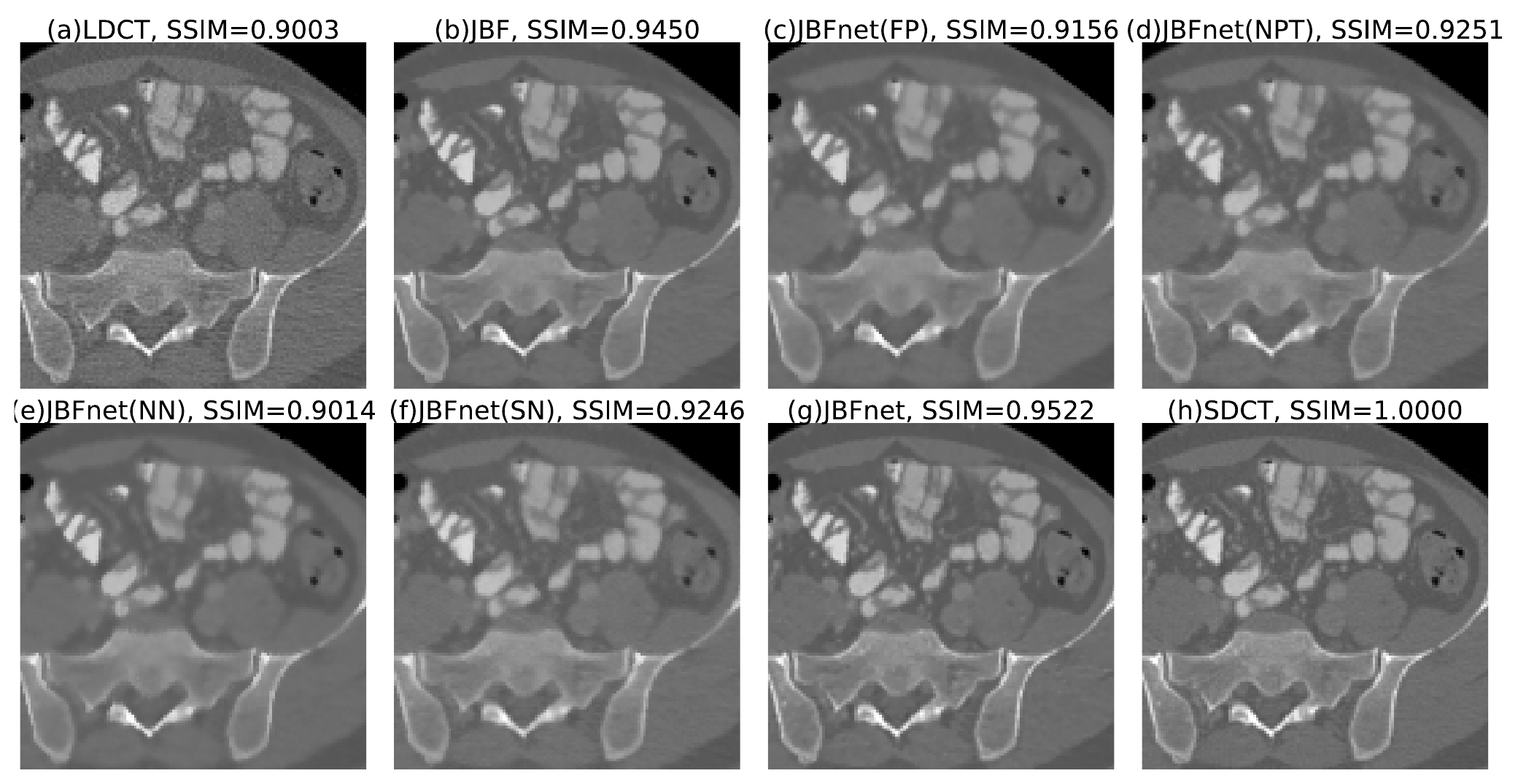}
	\caption{We display a close-up example of (a) low dose CT, denoised by (b) JBF (c) JBFnet with frozen prior (FP), (d) JBFnet with no pre-training (NPT), (e) JBFnet with no NM (NN) added (f) JBFnet with single-weight NM (SN) added (g) JBFnet, and finally we display (e) the standard dose CT. SSIM scores are displayed in the captions. JBFnet achieves the best performance (SSIM = 0.9522). Images are displayed with a window of [-800, 1200].}
	\label{abla}
\end{figure}

\subsection{Training and Implementation}
We trained JBFnet using the body scan data from 10 different patients. The scans were reconstructed using standard clinical doses, as well as 5\%, 10\%,  25\% and 50\% of the standard dose. The standard dose volumes are our reference volumes $I$. The reduced dose volumes are our noisy volumes $I_n$.

JBFnet was fed blocks of size 64 $\times$ 64 $\times$ 15. The guidance image $I_g$ was estimated from the input, which was shrunk to 64 $\times$ 64 $\times$ 7 by $P$. $I_g$ was then zero-padded to 70 $\times$ 70 $\times$ 7, and then shrunk to a 66 $\times$ 66 $\times$ 3 range kernel by $F$ (Fig. \ref{JBFnetComp}(c)). The distance matrix was constant across the whole image, and was pre-computed as a 7 $\times$ 7 $\times$ 7 matrix, which was shrunk to a 3 $\times$ 3 $\times$ 3 kernel by $G$. The noisy input $I_n$ to the JBF blocks was a 64 $\times$ 64 $\times$ 3 block, which contained the central 3 slices of the input block. $I_n$ was zero-padded to 66 $\times$ 66 $\times$ 3, and then shrunk to a 64 $\times$ 64 output by the bilateral filter (Fig. \ref{JBFnetComp}(c)). The NM was added in with learned pixelwise weights. After the JBF block, the output filtered image was padded by the neighbouring slices in  both directions, restoring the input size to 64 $\times$ 64 $\times$ 3 for future JBF blocks. This padding was not performed after the last JBF block, resulting in an overall output size of 64 $\times$ 64.

Training was performed for 30 epochs over the whole dataset. 32 image slabs were presented per batch. For the first ten epochs, only $W_p$ was updated ($\lambda_1$ = 0, $\lambda_2$ = 1, $\lambda_3$ = 0). This was to ensure a good quality guidance image for the JBF blocks. From epochs 10 - 30, all the weights were updated ($\lambda_1$ = 1, $\lambda_2$ = 0.1, $\lambda_3$ = 0.1). JBFnet was implemented in PyTorch on a PC with an Intel Xeon E5-2640 v4 CPU and an NVIDIA Titan Xp GPU.

\section{Experimental Results}
\begin{table}[t]
	\caption{Mean and standard deviations of the quality metric scores of our method and variants on the AAPM Grand Challenge dataset \cite{Mccollough2016}. The number of parameters in the networks are also included. Plain JBF application yields the highest PSNR ($46.79 \pm 0.8685$), while JBFnet yields the highest SSIM ($0.9825 \pm 0.0025$).}
	\begin{center}
		\begin{tabular}{||c||c c ||}
			\hline
			& PSNR & SSIM \\
			\hline
			Low Dose CT & $43.12 \pm 1.021$ & $0.9636 \pm 0.0073$ \\
			\hline
			CPCE3D \cite{Shan2018} & $45.43 \pm 0.6914$ & $0.9817 \pm 0.0029$ \\
			GAN\cite{Wolterink2017} & $41.87 \pm 0.7079$ & $0.9398 \pm 0.0079$ \\
			Deep GFnet \cite{Wu2018} & $41.62 \pm 0.3856$ & $0.9709 \pm 0.0029$ \\
			\hline
			JBF & $\boldsymbol{46.79 \pm 0.8685}$ & $0.9770 \pm 0.0046$ \\
			JBFnet (Frozen Prior)  & $42.86 \pm 0.4862$ & $0.9768 \pm 0.0027$ \\
			JBFnet (No pre-training)  & $42.75 \pm 0.5076$ & $0.9787 \pm 0.0027$ \\
			JBFnet (No NM)  & $42.64 \pm 0.5001$ & $0.9744 \pm 0.0029$ \\
			JBFnet (Single-weight NM)  & $42.07 \pm 0.5421$ & $0.9776 \pm 0.0029$ \\
			\hline
			\textbf{JBFnet} & $44.76 \pm 0.6009$ & $\boldsymbol{0.9825 \pm 0.0025}$ \\
			\hline
		\end{tabular}		
	\end{center}
	\label{resulttable}
\end{table}
\subsection{Test Data and Evaluation Metrics}
We tested our denoising method on the AAPM Low Dose CT Grand Challenge dataset \cite{Mccollough2016}. The dataset consisted of 10 body CT scans, each reconstructed at standard doses as used in the clinic and at 25\% of the standard dose. The slices were of 1mm thickness. We aimed to map the reduced dose images onto the standard dose images. The full dose images were treated as reference images. Inference was performed using overlapping 256 $\times$ 256 $\times$ 15 blocks extracted from the test data.

We used the PSNR and the SSIM to measure the performance of JBFnet. Since structural information preservation is more important in medical imaging, the SSIM is a far more important metric of CT image quality than the PSNR. Due to the small number of patients, we used the Wilcoxon signed test to measure statistical significance. A p-value of 0.05 was used as the threshold for determining statistical significance.

\subsection{Comparison to State-of-the-Art Methods}
We compare the denoising performance of JBFnet to other denoising methods that use deep learning. We compare JBFnet against CPCE3D \cite{Shan2018}, GAN \cite{Wolterink2017}, and deep GFnet \cite{Wu2018} (Fig. \ref{sota}). All networks were trained over 30 epochs on the same dataset. JBFnet achieves significantly higher scores in both PSNR and SSIM compared to GAN and  deep GFnet (w = 0.0, p = 0.005). CPCE3D achieves a significantly higher PSNR than JBFnet (w = 0.0, p = 0.005), but a significantly lower SSIM (w = 0.0, p = 0.005) (Table \ref{resulttable}). Additionally, the JBF block consists of only 112 parameters, compared to the guided filtering block \cite{Wu2018} which contains 1,555 parameters.

%\begin{table}[t]
%	\begin{center}
%		\begin{tabular}{||c||c||}
%			\hline
%			& Number of Parameters \\
%			\hline
%			CPCE3D \cite{Shan2018} & 118209 \\
%			GAN\cite{Wolterink2017} & 861857 (Gen.) + 513529 (Disc.) \\
%			JBFnet & 118209 (Prior est.) + 488 (JBF blocks) \\
%			\hline
%		\end{tabular}		
%	\end{center}
%	\caption{Number of parameters present in the networks. CPCE3D has the fewest number of parameters.}
%	\label{paramtable}
%\end{table}

\subsection{Ablation Study}

\subsubsection{Training the Filtering Functions}
Usually, the filter functions $F$ and $G$ of the bilateral filter are assumed to be Gaussian functions. We check if representing these functions with convolutions improves the denoising performance of our network. Training the filtering functions reduces our PSNR (w = 0.0, p = 0.005) but improves our SSIM (w = 0.0, p = 0.005) (Table \ref{resulttable} and Fig. \ref{abla}).

\subsubsection{Pre-training the Prior Estimator}
In our current training setup, we exclusively train the prior estimator $P$ for 10 epochs, to ensure a good quality prior image. We check if avoiding this pre-training, or freezing the value of $P$ after training improves the performance of our network. Both freezing $P$ and not doing any pre-training reduce the PSNR and SSIM (w = 0.0, p = 0.005) (Table \ref{resulttable} and Fig. \ref{abla}).

\subsubsection{Pixelwise Mixing of the NM}
Currently, we estimate the amount of the NM to be mixed back in by generating pixelwise coeffecients from a single 3 $\times$ 3 convolution of the NM. We check if not adding in the NM, or adding in the NM with a fixed weight improves the denoising performance of the network. Not mixing in the NM reduces our PSNR and SSIM significantly (w = 0.0, p = 0.005). Mixing in the NM with a fixed weight reduces both PSNR and SSIM even futher (w = 0.0, p = 0.005) (Table \ref{resulttable} and Fig. \ref{abla}).

\section{Conclusion}
In this study, we introduced JBFnet, a neural network which implements Joint Bilateral Filtering with learnable parameters. JBFnet significantly improves the denoising performance in low dose CT compared to standard Joint Bilateral Filtering. JBFnet also outperforms state-of-the-art deep denoising networks in terms of structural preservation. Furthermore, most of the parameters in JBFnet are present in the prior estimator. The actual filtering operations are divided into various JBF blocks, each of which has only 112 trainable parameters. This allows JBFnet to denoise while still maintaining physical interpretability.

\bibliographystyle{splncs04}
\bibliography{Bibliography}

\end{document}